\documentclass[preprint,showpacs,aps,prb]{revtex4}

\usepackage{graphicx}
\usepackage{dcolumn}
\usepackage{bm}

\begin{document}

\newcommand{\vektor}[1]{\mbox{\boldmath $#1$}}

\title{Magnetism in the dilute Kondo lattice model}

\author{M. Gulacsi, I. P. McCulloch}
\affiliation{Department of Theoretical Physics,
Institute of Advanced Studies, \\
The Australian National University,
Canberra, ACT 0200, Australia}

\author{A. Juozapavicius, A. Rosengren}
\affiliation{Department of Theoretical Physics,
Royal Institute of Technology,
SE-100 44 Stockholm, Sweden}
\date{April 14, 2003}

\begin{abstract}

The one dimensional dilute Kondo lattice model is investigated by means
of bosonization for different dilution patterns of the array of impurity
spins. The physical picture is very different if a commensurate or
incommensurate doping of the impurity spins is considered. For the
commensurate case, the obtained phase diagram is verified using a
non-Abelian density-matrix renormalization-group algorithm. The
paramagnetic phase widens at the expense of the ferromagnetic phase
as the $f$-spins are diluted. For the incommensurate case,
antiferromagnetism
is found at low doping, which distinguishes the dilute Kondo lattice
model from
the standard Kondo lattice model.

\end{abstract}


\maketitle

Heavy fermion systems have been of great theoretical interest
since their discovery some twenty years ago.\cite{hf} The central
problem posed by heavy fermion materials is to understand the
interaction between an array of localized moments (generally $f$ -
electrons in lanthanide or actinide ions) and conduction electrons
(generally $s$ - or $d$ - band). This situation is well described
by an antiferromagnetically coupled Kondo type model.

The solution of Kondo type models is well understood in two
limiting cases; the single impurity limit\cite{kondo}
which can be reduced to a one dimensional problem and solved
via Bethe Ansatz, and secondly the Kondo lattice model (KLM), which
was solved via bosonization\cite{boso,graeme} and numerous numerical
approaches\cite{ian,numerics} in one dimension for half - and partial
conduction band - filling. For half - filling the results indicate
the existence of a finite spin and charge gap. Accordingly in this
case the Kondo lattice model is an insulator with well defined
massive solitonic excitations of the spin sector.

For partial conduction band filling, the conduction electrons form a
Luttinger liquid, with spin and charge separation.\cite{boso,numerics}
The localized spins, however, exhibit ferromagnetism, due to an effective
double exchange coupling.\cite{graeme,ian} The double exchange is driving
the system toward ferromagnetism, while the fluctuations generated by
Kondo singlets compete against this tendency. As a consequence,
the paramagnetic to ferromagnetic phase transition is of the quantum
order - disorder type, typical to models with an effective
random field.\cite{graeme} However, for small Kondo coupling and close
to half filling an RKKY liquid state and polaronic regime are always
present.\cite{ian}

Beyond these two solvable limits, no rigorous results exist for
the intermediate cases, where the number of impurities are
neither one, nor equal to the number of sites. This is the
focus of our study. We concentrate on the one dimensional case,
and start from the Kondo lattice limit introducing impurity spin
{\sl holes}, that is we will be dealing with a {\sl dilute} Kondo
lattice model (DKLM):
\begin{equation}
H \: = \: -t \sum^{L-1}_{i = 1, \sigma} ( c^{\dagger}_{i, \sigma}
c^{}_{i + 1, \sigma} + {\rm h.c.} ) \: + \:
J \sum^{L}_{i = 1} {\cal P} \: {\bf S}^{c}_{i} {\bf \cdot}
{\bf S}^{}_{i} \: {\cal P} \; ,
\label{dklm}
\end{equation}
where $L$ is the number of sites and $t > 0$ is the conduction
electron hopping. We measure the Kondo coupling $J$ in units
of the hopping $t$. We denote by $N_{f}$ ($n_{f} = N_{f} / L$)
the number (concentration)  of impurities and $N_{c}$ ($n_{c} =
N_{c} / L$) the number (concentration) of conduction electrons.
The constraint $N_{f} \le L$ is imposed by ${\cal P}$, which
is an operator that projects out a pre-determined set of $f$-spins.
${\bf S}_{i}$ are spin $1/2$ operators for the
localized spins, e.g. $f$, and ${\bf S}^{c}_{i}= \frac{1}{2}
\sum_{\sigma^{},\sigma^{\prime}} c^{\dagger}_{i, \sigma^{}}
{\vektor{\sigma}}_{\sigma^{},\sigma^{\prime}} c^{}_{i, \sigma^{\prime}}$
with ${\vektor{\sigma}}$ the Pauli spin matrices and
$c^{\dagger}_{i, \sigma}$, $c^{}_{i, \sigma}$ the electron
creation and annihilation site operators.

We investigate the behavior of the DKLM both by an
analytical approach, based on a standard bosonization scheme,
and by numerical calculations. The latter were performed using
the newly developed non-Abelian density-matrix renormalization-group
(DMRG) algorithm,\cite{dmrg} which preserves the total spin and
pseudospin symmetry. This choice of basis first of all greatly
facilitates
the observation of magnetic phases and secondly it gives a dramatic
performance improvement compared to the standard DMRG basis.


The bosonization we use takes the standard approach\cite{haldane} by
first decomposing the on-site operators into Dirac fields, with spinor
components $\tau = \pm$ (otherwise known as the right, $\tau = +$, and
left, $\tau = -$, movers):
$c^{}_{x, \sigma} \: \approx \: \sum_{\tau} c^{}_{\tau, x, \sigma}
\: \equiv \: \sum_{\tau} e^{i k_{F} x} \Psi_{\tau, \sigma} (x)$,
where $k_{F} = \pi n_{c} /2$ and we consider the lattice spacing
to be unity. Next we bosonize the Dirac fields with
$\Psi_{\tau, \sigma} = \exp (i \Phi_{\tau, \sigma}) / \sqrt{2 \pi
\lambda}$,
where $1 / \lambda$ is the ultraviolet cutoff. For the scalar Bose
fields,
$\Phi_{\tau, \sigma} (x)$ and its canonical conjugate momenta,
$\Pi_{\tau, \sigma} (x)$, $\Phi_{\tau, \sigma} (x) = \int^{x}_{-\infty}
d x^{\prime} \Pi_{\tau, \sigma} (x^{\prime})$,
we use the standard Mandelstam representation,\cite{bosorep}
which introduces a momentum cutoff function
$\Lambda (k) = \exp ( - \lambda \vert k \vert / 2 )$ via the Fourier
transforms. Thus, the electron field can be represented in terms of
collective density operators which satisfy Bose commutation
relations:
\begin{equation}
c^{}_{\tau, x, \sigma} \: \approx \: \exp ( i \tau k_{F} x) \:
\exp i \{ \theta_{\rho}(x) + \tau \phi_{\rho}(x)
+ \sigma [ \theta_{\sigma}(x) + \tau \phi_{\sigma}(x)] \} / 2  \; ,
\label{two}
\end{equation}
where the Bose fields (for both $\nu = \rho, \sigma$) are defined by
$\phi_{\nu} / \theta_{\nu} = i (\pi / N) \sum_{k \ne 0} e^{i k x}
[ \nu_{+}(k) \pm \nu_{-}(k) ] \Lambda(k) / k$, where $\phi_{\nu}$
are the number fields and $\theta_{\nu}$ the current fields.
The charge (holon) and spin (spinon) number fluctuations are defined as
$\rho_{\tau}(k) = \sum_{\sigma} \rho_{\tau, \sigma}(k)$, and
$\sigma_{\tau}(k) = \sum_{\sigma} \sigma \rho_{\tau, \sigma}(k)$.
This type of Bose representation provides a non-perturbative description
of the conduction electrons in terms of\cite{haldane} holons and
spinons. We will neglect for the moment all the rapidly
oscillating (umklapp) terms. These will give a contribution
only at half filling, i.e. $n_{c} = n_{f}$, and will be analyzed
later on. Thus, the bosonized form of DKLM is:
\begin{eqnarray}
H  \: &=& \: \frac{v_{F}} {4 \pi} \sum_{j,\nu}
\left\{ \Pi_{\nu}^{2}(j) + [\partial_{x}\phi_{\nu}(j)]^{2} \right\}
\: + \: \frac{J}{2 \pi}\sum_{j} [\partial_{x}\phi_{\sigma}(j)] S_{j}^{z}
\nonumber \\
&+& \: \frac{J}{4 \pi \lambda} \sum_{j} \left\{
\cos [\phi_{\sigma}(j)] + \cos[2k_{F}j + \phi_{\rho}(j)] \right\}
\left(e^{-i \theta_{\sigma}(j)} S_{j}^{+} + {\rm h.c.} \right)
\nonumber \\
&-& \: \frac{J}{2 \pi \lambda} \sum_{j} \: \sin[\phi_{\sigma}(j)]
\sin[2k_{F}j + \phi_{\rho}(j)] S_{j}^{z} \; .
\label{bklm}
\end{eqnarray}
This equation has the same form as for a standard Kondo
lattice\cite{boso,graeme} except {\sl i)} we have to keep in
mind that the impurity spin, i.e. terms containing $S_{j}^{z}$,
$S_{j}^{+}$ and $S_{j}^{-}$, contributes only if there is an
$f$ spin at site $j$, and {\sl ii)} the even cutoff function
$\Lambda(k)$, defined in Eq.\ (\ref{two}), satisfying $\Lambda(k)
\approx 1$ for $\vert k \vert < 1/ \lambda$ and $\Lambda(k) \approx 0$
otherwise, is needed in the Bose fields to ensure that delocalized
conduction electrons are described. Delocalization is essential to
describe ferromagnetism,\cite{graeme,ian} since ferromagnetism
in the Kondo lattice models is due to the double exchange, which
only requires that $N_{c} < N_{f}$.\cite{zener}

In this situation each electron has on average more than one
localized spin to screen, and since hopping between localized
spins is energetically most favorable for electrons which preserve
their spin as they hop, this tends to align the underlying localized
spins.\cite{zener} This also means that double exchange will vanish
if the distance between impurity spins is larger than $\lambda$. At
lengths beyond $\lambda$ the electrons will behave as collective
density fluctuations, as usual in one dimensional systems.\cite{haldane}
Hence, $\lambda$ measures the effective range of the double exchange,
and in principle it is a function of $J$, $n_{c}$ and $n_{f}$.

The most straightforward method to determine an ordering of the
localized spins is by applying a unitary transformation ${\tilde{H}} =
e^{ {{\hat{\rm S}}} } H e^{- {{\hat{\rm S}}} }$. We choose
the transformation which changes to a basis of states in which
the conduction electron spin degrees of freedom are coupled directly
to the localized spins: ${{\hat{\rm S}}} = i \frac{J}{2 \pi v_{F}}
\sum_{j} \theta_{\sigma}(j) \: S_{j}^{z}$.  We perform the unitary
transformation up to infinite order, so there is not any
artificial truncation error generated (for details see Ref. 
\onlinecite{graeme}). In the new transformed basis the double
exchange interaction leading to ferromagnetism is clearly
exhibited and we obtain the effective Hamiltonian for the
localized spins:
\begin{eqnarray}
H_{{\rm eff}} \: &=& \: - {\frac{J^2}{2 \pi^2 v_{F}}}
\sum_{i, j} {\frac{\lambda}{\lambda^2 + (i - j)^2}} \:
S^{z}_{i} S^{z}_{j}
\nonumber \\
&+& \: {\frac{J}{2 \pi \lambda}} \sum_{i}
\{ \cos[K(i)] + \cos[2 k_{F} i] \} S^{x}_{i}
\nonumber \\
&-& \: {\frac{J}{2 \pi \lambda}} \sum_{i}
\sin[K(i)] \sin[2 k_{F} i] S^{z}_{i} \; .
\label{heff}
\end{eqnarray}
The $K(i)$ is a term introduced by the unitary transformation,
which counts all the $S^{z}_{i}$'s to the right of the site $i$
and subtracts from those to the left of $i$: $K(i) =
(J / 2 v_{F}) \sum_{l = 1}^{\infty} ( S^{z}_{i + l} -
S^{z}_{i - l} )$. This term gives the crucial difference between
KLM and DKLM, as will be explained later on.
The most important term in Eq.\ (\ref{heff}) is the first one,
which clearly shows that a ferromagnetic coupling emerges for DKLM.
This coupling is non-negligible for $N_{c} < N_{f}$ and
$i - j \le \lambda$ and its strength will decrease with increasing
distance between
impurity spins.

For $N_{c} < N_{f}$, the physical picture given by
Eq.\ (\ref{heff}) will be crucially different if the
lattice of impurity spins contains {\sl commensurate}
or {\sl incommensurate} array of holes. Hence, we analyze
these two cases separately. If we have a {\sl commensurate} doping
of the impurity spins, then we can approximate the ferromagnetic term in
the usual way by taking $\approx 1/ n_{f}$ for the shortest average
distance between $f$ spins: $\{ J^2 n^2_{f} \lambda / [2 \pi^2 v_{F}
(1 + n^2_{f} \lambda^2 ) ] \} \: \sum_{i} S^{z}_{i} S^{z}_{i + 1 /
n_{f}}$.
Lattice sites which are not occupied by $f$ spins are inert and
do not contribute to the ferromagnetic phase. This was verified by
DMRG: the calculated $f-f$ spin correlation functions behave
similarly as those of the normal KLM. The empty sites are inert.
The $f$-structure factor has the usual peak at $k/\pi = N_c/N_f$ for
low $J$, hence in the commensurate case the DKLM behaves similarly to the
standard KLM model.

To understand the behavior of the second and third term from Eq.\
(\ref{heff}), we notice that $K (i)$ is vanishingly small for the
commensurate case, as the number of $f$ impurity spins to the left and
to the right of a given site $i$ is the same. So the effective
Hamiltonian will reduce to the random transverse field Ising model, as
in the KLM.\cite{graeme} The randomness is generated by $(1 + \cos[2
k_{F} i])$ at large distances and it is driven by a cosine
distribution, similarly to spin-glasses.\cite{abrikosov} To determine
the phase transition we need the dependence $\lambda = \lambda (J,
n_{c}, n_{f})$, which however is very difficult to determine and as
such we use the low density value $\lambda \approx {\sqrt{2 / J}}$
close to criticality, similarly to previous works.\cite{graeme,shimoi}
In this way we obtain the critical phase transition line to be: $J =
\pi \sin( \pi n_{c} / 2) / [1 - \pi \sin( \pi n_{c} / 2) / n^2_{f}]$,
which represents a quantum order - disorder transition with variable
exponents.\cite{graeme} However, this ferromagnetic phase
disappears for larger distances between impurities because, as mentioned
earlier, the double exchange interaction vanishes if the average distance
between impurity spins, $1 / n_{f}$, is larger than $\lambda$. This is
very important because it ensures that the single impurity limit,
$n_{f} \rightarrow 0$, is free of ferromagnetism, as it should be.

The {\sl incommensurate} case is more difficult than the commensurate
case.
The reason is that in the low concentration limit the properties of
DKLM will be very much dependent on the random distribution of $f$
spins. We may observe phase separation or clusterization processes in
this case. In this limit, where the average distance between impurities
is very large, then the single impurity\cite{kondo} approximation
seems natural. However, if we look at small doping of $f$ electrons
only, then the main difference compared to the commensurate limit
studied previously is that the $K (i)$ term, in Eq.\ (\ref{heff}), is not
negligible anymore. The impurity $f$ spins are no longer equally
distributed to the left and right of a given site $i$, hence $K(i)
\approx (-1)^{i} (J / 2 v_{F})$, which gives rise to a staggered
field. The properties of Eq.\ (\ref{heff}) are then given by the
staggered field Ising model, which gives an antiferromagnetic ordering.
This antiferromagnetic ordering of the impurity spins represents a new
element in DKLM compared to the Kondo lattice. This corresponds to the
soliton lattice obtained by Schlottmann in a dynamical mean-field
treatment of the three dimensional dilute Kondo
lattice.\cite{schlottmann}

Similar behavior also occurs above half-filling, i.e. $N_{c} >
N_{f}$, where double exchange (as shown previously) does not
appear. But bosonization still works: the effective Hamiltonian
reduces to the second and third terms of Eq.\ (\ref{heff}), from which
the most dominant term, for low
doping of impurity spins, as in the case described previously,
is a staggered $S^{z}_{i}$ field. As the
first term in Eq.\ (\ref{heff}) is missing in this case, the only
fluctuation which
can destroy a locked staggered order is $S^{x}_{i}$. For large $J$ ($
4 \lesssim J$) the staggered order wins. While for smaller values of
$J$ the systems will be disordered.

As we approach half filling from both sides, the bosonization approach
breaks down as the strongly oscillating (umklapp) fields start
dominating. The DKLM will undergo a metal-insulator transition as in a
standard quantum sine-Gordon model\cite{oldpaper} by dynamical mass
generation. A Spin gap will also appear. This can be understood easily,
because the half filled DKLM is equivalent to the quarter filled
periodic Anderson model, which has a antiferromagnetic
order.\cite{shimoi} The only difference from the Kondo lattice is that
the
massive solitons obtained for DKLM are of Su, Schrieffer and Heeger
type.\cite{SSH}


To confirm the previously obtained magnetic phases, we performed
non-Abelian DMRG analysis of the DKLM model in the commensurate
case, for both the $N_{c} < N_{f}$ and $N_{c} > N_{f}$,
where ferro- and antiferromagnetism, respectively exist.
For the $N_{c} < N_{f}$ case, we have investigated several
commensurate dilution patterns of the form:
``0 0 $f$ 0 0 $f$ 0 0 $f$ $\ldots$'';
``0 $f$ 0 $f$ 0 $f$ 0 $f$ $\ldots$''; and
``0 $f$ $f$ 0 $f$ $f$ 0 $f$ $f$ $\ldots$''
in a  64-site long chain. While in the $N_{c} > N_{f}$ limit we have
studied the $n_{f} = 0.8$ localized spin filling (to be in accordance
with the bosonization requirement of low dilution) on a 80-site long
DKLM chain, i.e., we investigated the pattern:
``0 $f$ $f$ $f$ $f$ 0 $f$ $f$ $f$ $f$ $f$ $\ldots$''.

These patterns were selected in such a way that the chain middle-point
reflection symmetry was preserved to accelerate the calculations. There
was only one exception: the ``0 $f$ 0 $f$ 0 $f$'' chain has an impurity
in the middle: ``0 $f$ 0 $f$ $f$ 0 $f$ 0'', but its effect is rather
small compared to our final errors and so it was neglected.

In the DMRG calculations careful error and convergence analysis were
used, and we extrapolated the energy linearly to zero truncation
error (we saw no quadratic terms large enough to affect the fit).
For each given dilution pattern, filling $N_c/N_f$, and
interaction constant $J$, we used several DMRG sweeps of between
200 and 400 $SO(4)$-symmetric states.

\begin{figure}
\includegraphics[width=12cm]{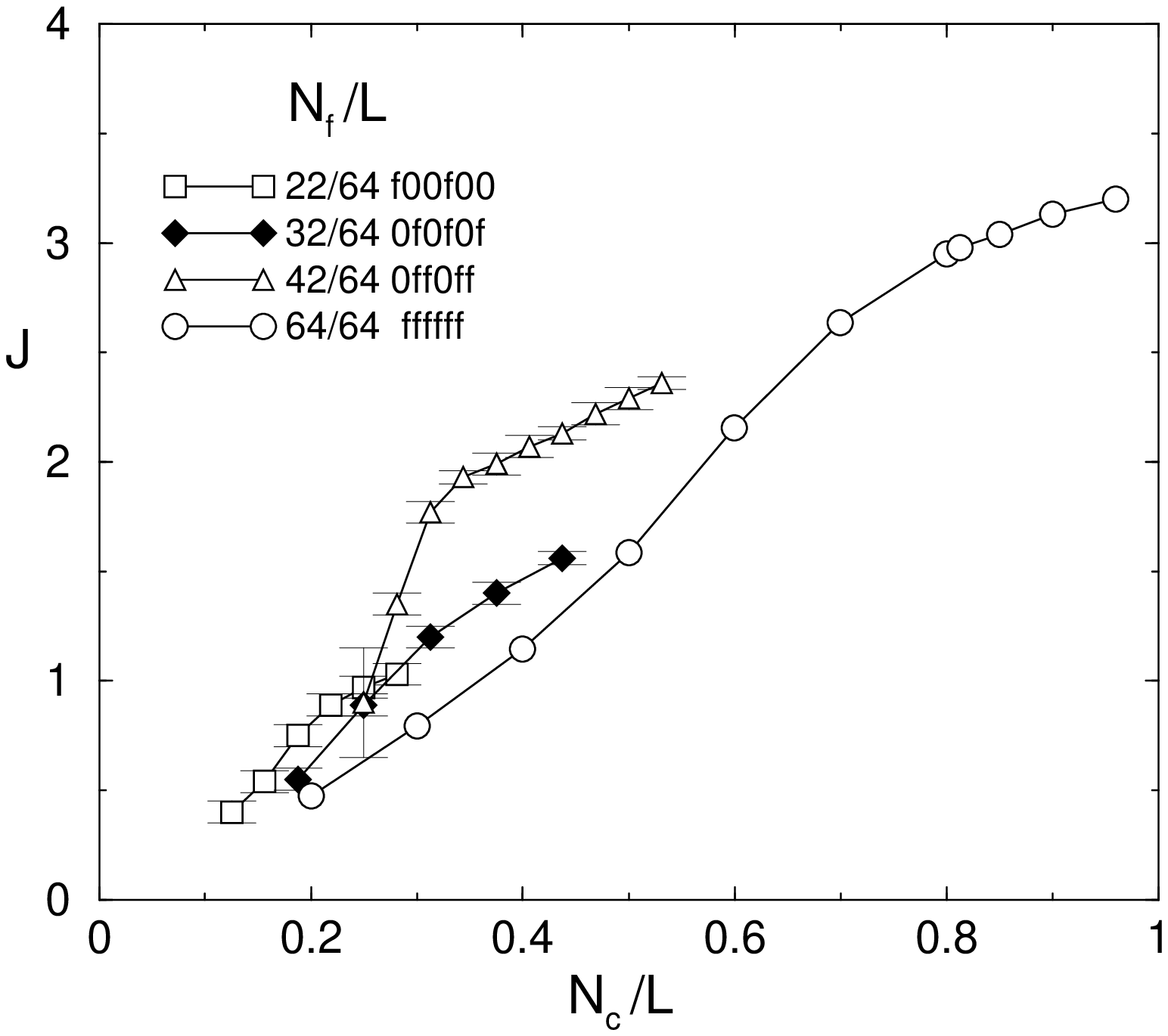}
\caption{\label{fig:Fig1} The phase diagram of the dilute Kondo model
in different commensurate filling cases for $N_{c} < N_{f}$. Legend
shows patterns of dilution. Open circles correspond to the standard
KLM model. The system of a given dilution pattern is ferromagnetic
above the corresponding solid line and paramagnetic below.}
\end{figure}

For $N_{c} < N_{f}$ ferromagnetism appears at large $J$ values,
see Fig.\ \ref{fig:Fig1}. A point on the phase diagram shown
in Fig.\ \ref{fig:Fig1} is judged to be ferromagnetic if the extrapolated
energy of the spin $S_{\rm{max}}$ run is lower than the spin zero energy.
This energy difference to the spin singlet excited state can be
calculated
directly using the $SO(4)$ basis set of the non-Abelian DMRG.
The phase transition line can be determined with high accuracy if one
plots the energy gap between the spin $S_{\rm{max}}$ and spin zero
states as a function of $J$ - the gap rapidly decreases as the transition
is approached. It should be mentioned that, as well as in the standard
Kondo case,\cite{ian} there also exists additional ferromagnetic phases
inside
the paramagnetic region for dilute Kondo chains.

\begin{figure}
\includegraphics[width=12cm]{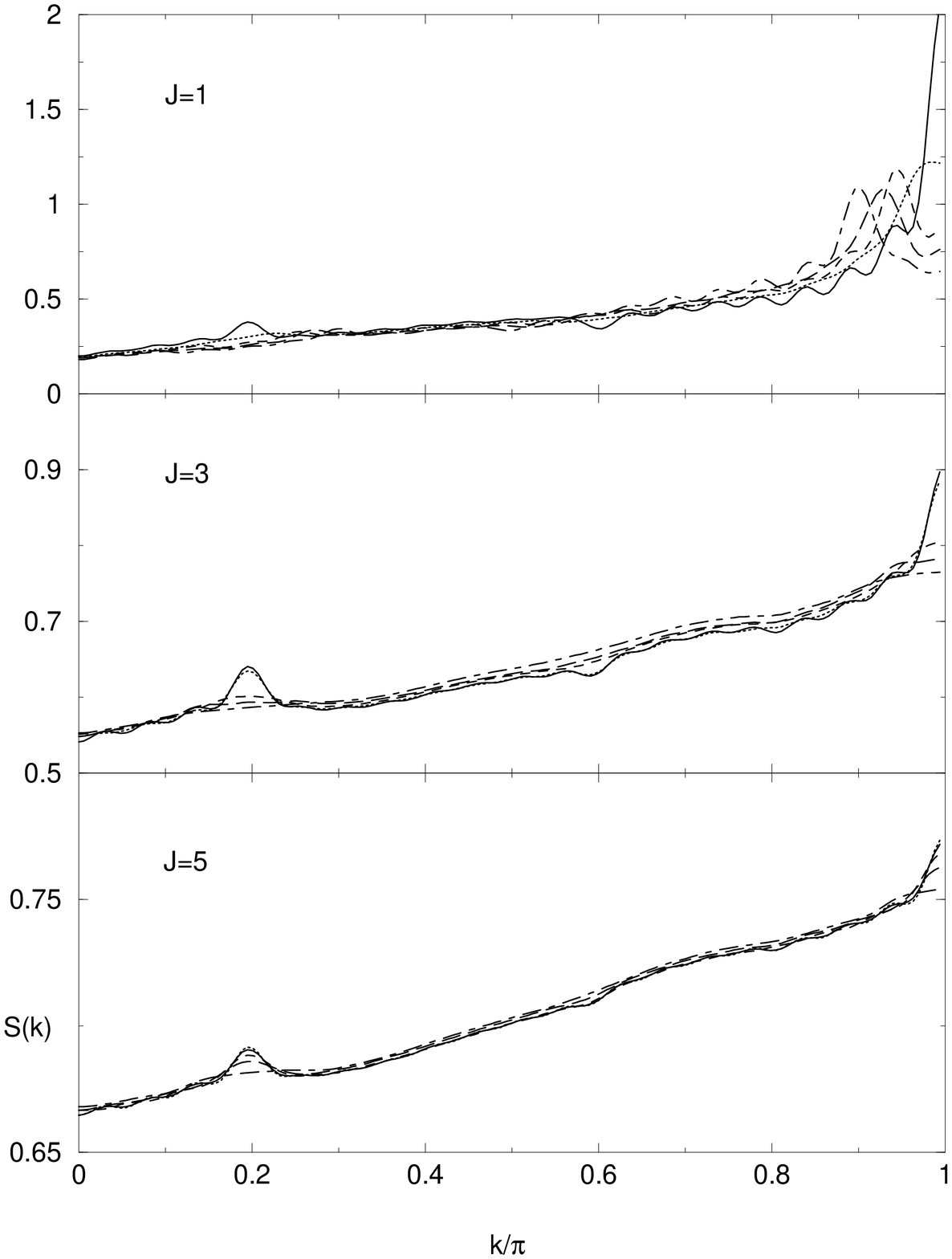}
\caption{\label{fig:Fig2} Typical $J$ dependences of the conduction
electron spin structure factor, $S(k)$ for $n_{f} = 0.8$ and
$n_{c} = 1$ continuous, $n_{c} = 0.975$ dotted, $n_{c} = 0.95$ dashed,
$n_{c} = 0.925$ long dashed and $n_{c} = 0.90$ dot-dashed curves,
respectively.}
\end{figure}

In the opposite limit, i.e., $N_{c} > N_{f}$, we have confirmed
the existence of antiferromagnetism by calculating the spin
structure factor $S (k)$. As can be seen in Fig.\ \ref{fig:Fig2},
for small $J$ and away from half filling the peak in $S(k)$ is at
$2 k_{F}$ (similarly to KLM \cite{ian}), i.e., the DKLM is a disordered
paramagnet. The dominant $2 k_{F}$ backscattering processes are manifest
of a system of free localized spins embedded in effective fields
determined
by conduction electron scattering. The half filled case, i.e., $n_{c} =
1$
as in the KLM model \cite{halffilled} is always an antiferromagentic
insulator.
The tendency of Kondo lattice models to form an antiferromagnetic
insulator
at half filling for any nonzero $J$ is a consequence of the Kondo
effect.\cite{halffilled} As we increase $J$, the distinctive
antiferromagnetic peak at $k = \pi$ appears away from half filling
(see Fig.\ \ref{fig:Fig2}). As we further increase $J$ the $2 k_{F}$ peak
broadens and decreases in strength while the antiferromagentism becomes
more robust, at $J = 5$ it appers already at $n_{c} \approx 0.925$.


In conclusion, we have studied the dilute Kondo lattice model in one
dimension both numerically, using DMRG, and analytically,
with a standard bosonization approach. We have derived an effective
Hamiltonian for the $f$ spins, which accounts for the appearance of a
ferromagnetic phase seen with a commensurate dilution pattern of the
impurity
spin array. The paramagnetic-ferromagnetic phase transition shifts to
higher coupling $J$ values as the $f$-spins of the chain are diluted,
in agreement with the numerical DMRG calculation. We have also shown
that for incommensurate dilution or above half filling, i.e.,
$N_{c} > N_{f}$ for low doping of impurity spins, antiferromagnetism
is found. This distinguishes the dilute Kondo lattice model from the
standard Kondo lattice model.

Work in Australia was supported by the Australian Research Council
and Department of Industry, Science and Resources. In Sweden
by The Swedish Natural Science Research Council, The Swedish
Foundation for International Cooperation in Research and Higher
Education (STINT), and The G\"{o}ran Gustafsson foundation.
Some of the numerical calculations were performed at
National Facility of the Australian Partnership for Advanced Computing,
via an award from the ANU Supercomputer Time Allocation Committee.

\newpage

\begin{thebibliography}{99}

\bibitem{hf}See, for example, P. A. Lee, T. M. Rice,
J. W. Serene, L. J. Sham, and J. W. Wilkins, Comments
Condens. Matter Phys. {\bf 12}, 99 (1986); P. Fulde,
J. Keller and G. Zwicknagl, in {\sl Solid state Physics:
Advances in Research and Applications}, Academic Press,
New York, 1988, vol. 41, p. 1

\bibitem{kondo}P. W. Anderson, G. Yuval and D. R. Hamann,
Phys. Rev. B{\bf 1}, 2719 (1989); N. Andrei, Phys. Rev. Lett.
{\bf 45}, 379 (1980); N. Andrei, K. Furuya and J. H. Lowenstein,
Rev. Mod. Phys. {\bf 55}, 331 (1983).

\bibitem{boso}O. Zachar, S. A. Kivelson, and V. J. Emery,
Phys. Rev. Lett. {\bf 77}, 1342 (1996); S. Fujimoto and N. Kawakami,
J. Phys. Soc. Japan {\bf 63}, 4322 (1994).

\bibitem{graeme}G. Honner and M. Gulacsi, Phys. Rev. Lett.
{\bf 78}, 2180 (1997); Phys. Rev. B{\bf 58}, 2662 (1998).

\bibitem{ian}I. P. McCulloch, M. Gulacsi, S. Caprara,  
A. Juozapavicius and A. Rosengren, J. Low Temp. Phys.
{\bf 117}, 323 (1999); I. P. McCulloch, A. Juozapavicius,
A. Rosengren and M. Gulacsi, Phil. Mag. Lett. {\bf 81},
869 (2001); and Phys. Rev. B{\bf 65}, 52410 (2002). 

\bibitem{numerics}M. Troyer and D. W\"{u}rtz, Phys. Rev.
B{\bf 47}, 2886 (1993); H. Tsunetsugu, M. Sigrist, and K. Ueda,
Phys. Rev. B{\bf 47}, 8345 (1993); S. Moukouri and L. G. Caron,
Phys. Rev. B{\bf 52}, R15723 (1995); M. Guerrero and C. C. Yu
Phys. Rev. B 51, 10301 (1995); C. C. Yu and M. Guerrero
Phys. Rev. B 54, 8556 (1996); S. Caprara and A. Rosengren,
Europhys. Lett. {\bf 39}, 55 (1997).

\bibitem{dmrg}I. P. McCulloch and M. Gulacsi, Aust. J. Phys.
{\bf 53}, 597 (2000); Phil. Mag. Lett. {\bf 81}, 447 (2001);
and Europhys. Lett. {\bf 57}, 852 (2002).

\bibitem{haldane}F. D. M. Haldane, J. Phys. C{\bf 14}, 2585 (1981).
For a review, see J. Voit, Rep. Prog. Phys. {\bf 57}, 977 (1994);
M. Gulacsi, Phil. Mag. B{\bf 76}, 731 (1997); J. von Delft and H.
Sch{\"o}ller, Ann. Phys. {\bf 4}, 225 (1998).

\bibitem{bosorep}S. Mandelstam, Phys. Rev. D{\bf 11}, 3026 (1975);
T. Banks, D. Horn and H. Neuberger, Nucl. Phys. B{\bf 108}, 199 (1976),
see also Y. K. Ha, Phys. Rev. D{\bf 29}, 1744 (1984).

\bibitem{zener}C. Zener, Phys. Rev. {\bf 82}, 403 (1951); P. W.
Anderson and H. Hasegawa, Phys. Rev. {\bf 100}, 675 (1955).

\bibitem{abrikosov}A. A. Abrikosov, Adv. Phys. {\bf 29}, 869 (1980).

\bibitem{shimoi}T. Yanagisawa and M. Shimoi, Int. J. Mod. Phys.
B{\bf 10}, 3383 (1996).

\bibitem{schlottmann}P. Schlottmann, Phys. Rev. B{\bf 46}, 998 (1992).

\bibitem{oldpaper}M. Gulacsi and K. S. Bedell,
Phys. Rev. Lett. 72, 2765 (1994).

\bibitem{SSH}S. W. Su, J. R. Schrieffer and A. J. Heeger,
Phys. Rev. B{\bf 22}, 2099 (1980).

\bibitem{footnote}
Note that each block state of our non-abelian DMRG corresponds to several
ordinary DMRG states, since it is a spin multiplet rather than
an individual state (see Ref.\ \protect\onlinecite{dmrg}).

\bibitem{halffilled}R. Jullien and P. Pfeuty, J. Phys. F{\bf 11},
353 (1981); H. Tsunetsugu, Y. Hatsugai, K. Ueda and M. Sigrist,
Phys. Rev. B{\bf 46}, 3175 (1992); C. C. Yu and S. R. White, Phys. Rev.
Lett. {\bf 71}, 3866 (1993); M. Guerreri and C. C. Yu, Phys. Rev. B{\bf
51},
10301 (1995).

\end{thebibliography}
\end{document}